\documentclass[useAMS,usenatbib]{mn2e}

\usepackage{graphicx,amsmath,amssymb}

\newcommand{\dif}{\mathrm{d}}
\newcommand{\map}{M_\mathrm{ap}}

\newcommand{\ex}{\mathrm{e}}
\newcommand{\h}{h_{70}}
\newcommand{\hm}{h_{70}^{-1}}

\renewcommand{\vec}[1]{\mbox{\boldmath$#1$}}

\usepackage[totalwidth=480pt, totalheight=680pt]{geometry}

\title{Cosmology with the shear-peak statistics} 
\author[J.\,P. Dietrich and
J. Hartlap]{J.\,P. Dietrich$^{1}$\thanks{E-mail: jorgd@umich.edu
    (JPD); hartlap@astro.uni-bonn.de (JH)} and J. Hartlap$^{2}$\\
$^{1}$ESO, Karl-Schwarzschild-Str. 2, 85748 Garching
  b. M\"unchen, Germany\\
$^{2}$Argelander-Institut f\"ur Astronomie, Auf dem H\"ugel 71, 53121 Bonn,
Germany}

\begin{document}

\date{Accepted 2009 October 23.  Received 2009 September 26; in
  original form 2009 June 16}

\pagerange{\pageref{firstpage}--\pageref{lastpage}} \pubyear{2009}
\maketitle

\begin{abstract}
  Weak-lensing searches for galaxy clusters are plagued by low
  completeness and purity, severely limiting their usefulness for
  constraining cosmological parameters with the cluster mass function.
  A significant fraction of `false positives' are due to projection of
  large-scale structure and as such carry information about the matter
  distribution.
  We demonstrate that by constructing a ``peak function'', in analogy
  to the cluster mass function, cosmological parameters can be
  constrained.
  To this end we carried out a large number of cosmological $N$-body
  simulations in the $\Omega_\mathrm{m}$-$\sigma_8$ plane to study the
  variation of this peak function.
  We demonstrate that the peak statistics is able to provide
  constraints competitive with those obtained from cosmic-shear
  tomography from the same data set. By taking the full
  cross-covariance between the peak statistics and cosmic shear into
  account, we show that the combination of both methods leads to
  tighter constraints than either method alone can provide.
\end{abstract}


\maketitle

\label{firstpage}

\section{Introduction}
\label{sec:introduction}
The number density of clusters of galaxies is a sensitive probe for
the total matter density of the Universe $\Omega_\mathrm{m}$, the
normalisation of the power spectrum $\sigma_8$, and the evolution of
the equation of state of the Dark Energy $w$ \citep[e.g.,][]{1998ApJ...508..483W,2001ApJ...553..545H,2002PhRvL..88w1301W}. For
some time it was thought that weak gravitational lensing, which by its
nature is sensitive to dark and baryonic matter alike and independent
of the dynamical or evolutionary state of the cluster, could be used
to construct clean, purely mass selected cluster samples. However,
ray-tracing simulations through cosmological $N$-body simulation made
it clear that weak-lensing selected clusters are not at all mass
selected but selected by the shear of the projected mass along the
line of sight
\citep[e.g.,][]{2004MNRAS.350..893H,2005ApJ...624...59H,2007A&A...470..821D}.
As a result, blind searches for galaxy clusters using weak lensing
have both low purity and completeness
\citep[e.g.,][]{2007A&A...462..875S,2007A&A...470..821D}.

Gravitational lensing is, due to the large intrinsic ellipticity
scatter of background galaxies, an inherently noisy technique. This
shape noise is the dominant noise source at the low signal-to-noise
ratio (SNR) end of the weak-lensing selection function, while
projections of large-scale structure (LSS) along the line-of-sight
(LOS) dominate the noise budget of highly significantly detected peaks
\citep{2007A&A...470..821D}. Both sources of noise affect purity and
completeness. Galaxy clusters aligned with underdense regions are not
visible as significant overdensities, while the projection of
uncorrelated overdensities can mimic the shear signal of galaxy
clusters. While these effects can be taken into account
\citep{2006PhRvD..73l3525M}, they degrade the constraints on
cosmological parameters one can obtain using weak-lensing selected
galaxy clusters.

Of course such projected peaks are noise or false positives only in
the sense of galaxy cluster searches. They are caused by real
structures along the line-of-sight and as such carry information about
the matter power spectrum. Whereas analytical models exist for the
halo mass function \citep{1974ApJ...187..425P,2002MNRAS.329...61S}, no
such model exists for the number density of peaks in weak lensing
surveys. Probably no such prediction can be made analytically because
the abundance of peaks depends on projections of uncollapsed yet
highly non-linear structures like filaments of the cosmic web. As an
additional complication the observed number of peaks depends on
observational parameters like limiting magnitude, redshift
distribution, and intrinsic ellipticity dispersion.

In the absence of an analytic framework, ray-tracing through $N$-body
simulations can be used to numerically compute the ``peak function''
(in analogy to the mass function) for a survey and study its variation
with cosmological parameters. Here we present a large set of such
simulations aimed at demonstrating the usefulness of the shear-peak
statistics for constraining cosmological parameters. We consider this
work to be a pilot study and limit ourselves to the variation of the
peak function with $\Omega_\mathrm{m}$ and $\sigma_8$ and its ability
to break the degeneracy between these two parameters encountered in
the 2-point cosmic-shear correlation-function. Unlike
\citet{2009ApJ...698L..33M} who showed that the projected mass
function, which is difficult to measure, scales with cosmology
essentially in the same way as the halo mass function, we study the
cosmological dependence of the directly observable aperture mass
statistics.

\section{Methods}
\label{sec:methods}
\subsection{$N$-body simulations}
\label{sec:n-body-simulations}
\begin{figure}
  \resizebox{\hsize}{!}{\includegraphics[clip]{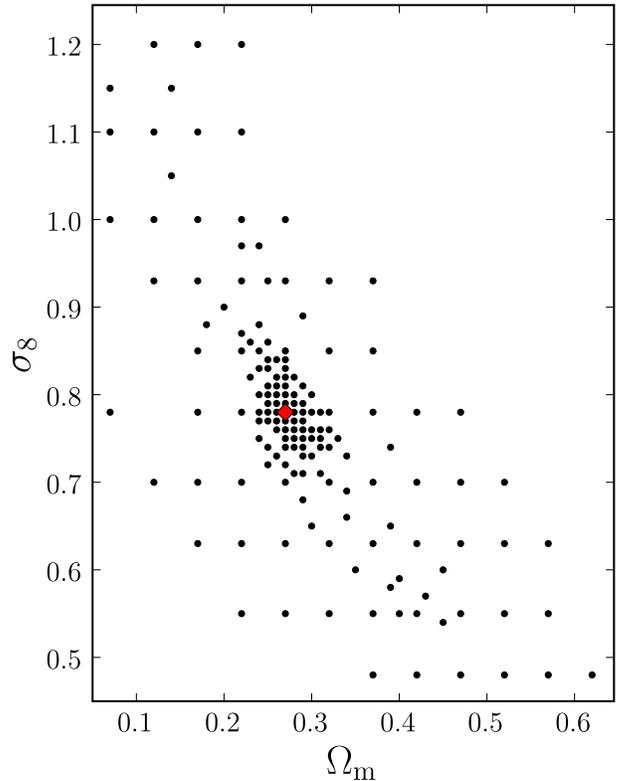}}
  \caption{Location of the 158 different cosmologies in the
    $\Omega_\mathrm{m}$-$\sigma_8$ plane for which $N$-body simulations were
    computed. The red diamond marks the fiducial cosmology at
    $(\Omega_\mathrm{M}, \sigma_8) = (0.27, 0.78)$.}
  \label{fig:simulations}
\end{figure}

We carried out $N$-body simulations for $158$ different flat
$\Lambda$CDM cosmologies with varying $\Omega_\mathrm{m}$,
$\Omega_\Lambda$, and $\sigma_8$. Figure~\ref{fig:simulations} shows
the distribution of these simulations in the
$\Omega_\mathrm{m}$-$\sigma_8$ plane. All simulations had $256^3$ dark
matter particles in a box with $200\,\hm$\,Mpc side length. These
choices reflect a compromise we had to make between computing a large
number of simulations to sample our parameter space on the one hand
and to have a fair representation of very massive galaxy clusters
dominating the cosmological sensitivity of the halo mass function on
the other hand. These simulation parameters were chosen such that we
can expect the presence of $10^{15}\,\hm\,M_\odot$ mass halos at
redshift $z = 0$ in the simulation box in our choice of fiducial
cosmology $\vec{\pi_0} = (\Omega_\mathrm{m_0} = 0.27, \Omega_\Lambda = 0.73,
\Omega_\mathrm{b} = 0.04, \sigma_{8_0} = 0.78, n_\mathrm{s} =
1.0, \Gamma = 0.21, \h = 1)$.

We computed $35$ $N$-body simulations for this fiducial cosmology to
estimate the covariance of our observables. The total number of
$N$-body simulations is thus $192$. Particle masses depend on the
cosmology and range from $m_\mathrm{p} = 9.3 \times 10^9\,M_\odot$ for
$\Omega_\mathrm{m} = 0.07$ to $m_\mathrm{p} = 8.2 \times
10^{10}\,M_\odot$ for $\Omega_\mathrm{m} = 0.62$. The particle mass at
our fiducial cosmology is $m_\mathrm{p} = 3.6 \times
10^{10}\,M_\odot$.

The $N$-body simulations were carried out with the publicly available
TreePM code GADGET-2 \citep{2005MNRAS.364.1105S}. The initial
conditions were generated using the \citet{1998ApJ...496..605E}
transfer function. We started the simulations at $z = 50$ and saved
snapshots in $\Delta z$ intervals corresponding to integer multiples
of the box size, such that we have a suitable snapshot for each lens
plane of the ray-tracing algorithm. The Plummer-equivalent force
softening length was set to $25\,\hm$\,kpc comoving. We checked the
accuracy of our $N$-body simulations by comparing their matter power
spectra with the fitting formula of \citet{2003MNRAS.341.1311S}.
Additionally, we also detected halos using a friend-of-friend halo
finder and compared their mass function to that of
\citet{2001MNRAS.321..372J}. All tests were done for a number of
different cosmologies and redshifts to ensure that the simulations
match our expectations over the parameter and redshift range under
investigation here.

\subsection{Ray-tracing}
\label{sec:ray-tracing}

We used the multiple lens-plane algorithm
\citep[e.g.][]{1986ApJ...310..568B,1992grle.book.....S,1994CQGra..11.2345S,2000ApJ...530..547J,2009A&A...499...31H}
to simulate the propagation of light rays through the matter
distribution provided by the $N$-body simulations: for a given
$N$-body simulation, we constructed the matter distribution along the
line of sight by tiling snapshots of increasing redshift. The matter
distribution of each snapshot was projected onto a lens plane located
at the snapshot redshift.

Note that the boxes are just small enough for the cosmic evolution
during the light travel time through a box to be negligible.  This
ensures that the matter distribution does not change significantly in
the volume that is represented by a particular lens plane, and that
the scale factor and the comoving angular diameter distances to the
structure projected onto this plane are essentially the same. If the
latter were not the case, this would lead to an erroneous conversion
of physical scales on the lens plane to angular scales on the sky. Our
allowed us to project a complete snapshot onto one lens plane instead
of creating several smaller redshift slices as was done in
\citet{2009A&A...499...31H}, which reduces the complexity of the
ray-tracing considerably.

Since the snapshots basically contain the same matter distribution at
slightly different stages of evolution, measures have to be taken to
avoid the repetition of structures along the line of sight. Making use
of the periodic boundary conditions of the simulation volume, we
applied random rotations, translations, and parity flips to the matter
distribution of each snapshot prior to the projection. For the
ray-tracing, we assumed that light rays are only deflected at the lens
planes and propagate freely in between. We compute the Fourier
transform of the deflection potential on each plane from the projected
mass density by solving the Poisson equation in Fourier space using
FFT, again exploiting the periodic boundary conditions.  From this,
the Fourier transforms of the deflection angles and their derivatives
can be obtained using simple multiplications. Finally, these
quantities are transformed to real space using an inverse FFT.  More
details on the formalism can be found, e.g., in
\citet{2000ApJ...530..547J}. With this, a set of light rays (forming a
grid in the image plane) can be propagated from the observer through
the array of lens planes using a recursion formula
\citep[see][]{2009A&A...499...31H}. Similarly, the Jacobian matrix of
the lens mapping from the observer to each of the lens planes can be
obtained.

We then sampled the image plane uniformly with galaxies, the redshift
of which was drawn from a distribution of the form
\begin{equation}
  \label{eq:1}
  p(z) \propto
    \left(\frac{z}{z_0} \right)^\alpha \exp\left[ -\left(
        \frac{z}{z_0} \right)^\beta \right]\;.
\end{equation}
The Jacobian matrices were interpolated from the grid onto the
galaxies (in the plane of the sky as well as in redshift) and the
reduced shear was computed. We simulated a CFHTLS-Wide like survey
for which we created five $6\times 6$\,sq.\,deg. patches from every
$N$-body simulation. The parameters of the redshift distribution
(\ref{eq:1}) were set to $\alpha = 0.836$, $\beta = 3.425$, and $z_0 =
1.171$, as determined for the CFHTLS-Wide \citep{2007MNRAS.381..702B}.
We set the galaxy number-density to $n_\mathrm{g} = 25$\,arcmin$^{-2}$
and the intrinsic ellipticity dispersion to $\sigma_\varepsilon =
0.38$. Because very few galaxies are present at high redshifts, the
redshift distribution was cut off at $z=3.0$ to save computing time.

\subsection{Peak detection}
\label{sec:peak-statistics}
\subsubsection{Aperture mass in 2-d}
\label{sec:aperture-mass-2}
The tidal gravitational field of matter along the line-of-sight causes
the shear field $\gamma(\vec{\theta})$ to be tangentially aligned
around projected mass-density peaks. We can use this tangential
alignment directly to detect weak-lensing peaks, instead of searching
for convergence peaks on maps of reconstructed surface mass-density as
it has been done sometimes in weak-lensing cluster searches
\citep[e.g.,][]{2007A&A...462..459G,2007ApJ...669..714M}. We define
the aperture mass \citep{1996MNRAS.283..837S} at position
$\vec{\theta}_0$ to be the weighted integral
\begin{equation}
  \label{eq:2}
  \map(\vec{\theta}_0) = \int\limits_{\mathrm{sup} Q} \dif^2\theta\,
  Q(\vartheta)\gamma_\mathrm{t}(\vec{\theta};\vec{\theta}_0)
\end{equation}
over the shear component tangential to the line $\vec{\theta}_0 -
\vec{\theta}$, $\gamma_\mathrm{t}(\vec{\theta}; \vec{\theta}_0)$. Here
$Q(\vartheta) = Q(|\vec{\theta}|)$ is a radially symmetric, finite and
continuous weighting function with $\lim_{\vartheta \rightarrow
  \infty} Q(\vartheta) = 0$. For later convenience we also require $Q$
to be normalised to unit area. If $Q(\vartheta)$ follows the expected
shear profile of a mass peak, the aperture mass becomes a matched
filter technique for detecting such mass peaks. On data the shear
field is sampled by galaxies with ellipticities $\varepsilon_i$. Then
$\map$ can be estimated by the sum over $N_\mathrm{g}$ galaxies in the
aperture,
\begin{equation}
  \label{eq:3}
  \hat{M}_\mathrm{ap} = \frac{1}{n_\mathrm{g}}\sum\limits_{i = 1}^{N_\mathrm{g}}
  Q(\vartheta_i) \varepsilon_{i\mathrm{t}}\;,
\end{equation}
where $\varepsilon_{i\mathrm{t}}$ is the tangential ellipticity
component of the $i$-th galaxy, defined in analogy to
$\gamma_\mathrm{t}$ above.

The SNR of the aperture mass can be computed directly from the data,
making use of the fact $\langle \map \rangle \equiv 0$. Then the RMS
dispersion is $\sigma_{\map} = \sqrt{\langle \map \rangle^2}$, which
can be estimated by
\begin{equation}
  \label{eq:4}
  \hat\sigma_{\map} = \frac{\sigma_\varepsilon}{\sqrt{2}n_\mathrm{g}}
  \left[\sum\limits_{i = 1}^{N_\mathrm{g}} Q^2(\vartheta_i)
    \right]^{1/2}\;,
\end{equation}
where we have made use of the fact that
\begin{equation}
  \label{eq:5}
  \langle \varepsilon_i \varepsilon_j \rangle =
  \frac{\sigma_\varepsilon^2}{2} \delta_{ij}\;,
\end{equation}
with $\sigma_\varepsilon$ being the intrinsic ellipticity
dispersion. The estimator for the SNR of the aperture mass is then
finally
\begin{equation}
  \label{eq:6}
  \hat S(\vec{\theta}_0) = \frac{\sqrt{2} \sum_i Q(\vartheta_i)
    \varepsilon_{i\mathrm{t}}}{\sqrt{ \sum_i Q^2(\vartheta_i)
      \varepsilon_i^2}}\;. 
\end{equation}

\subsubsection{Tomographic aperture mass}
\label{sec:tomogr-apert-mass}
The aperture mass statistics locates convergence peaks only in
projection on the sky. Using redshift information on the background
galaxies, e.g., from photometric redshifts, one can generalise the
2-dimensional aperture mass to a tomographic measure that is able to
deproject structures along the line-of-sight and locate peaks in
redshift space \citep{2005ApJ...624...59H}. The likelihood that a peak
at a position $\vec{\theta}_0$ is at a redshift $z_\mathrm{d}$ is
given by
\begin{equation}
  \label{eq:7}
  \ln \mathcal{L}(\vec{\theta}_0, z_\mathrm{d}) =
  \frac{1}{\sigma^2_\varepsilon} 
  \frac{\left[ \sum_i^{n_\mathrm{z}} Z(z_i; z_\mathrm{d})
      \map(\vec{\theta}_0) \right]^2}{\sum_i^{n_\mathrm{z}} Z^2(z_i;
    z_\mathrm{d})} \;, 
\end{equation}
where $Z(z_i; z_\mathrm{d})$ is the redshift weight for a background
galaxy in the $i$th redshift bin,
\begin{equation}
  \label{eq:8}
  Z(z; z_\mathrm{d}) = \frac{D_\mathrm{d} D_\mathrm{ds}}{D_\mathrm{s}}
  \mathcal{H}(z - z_\mathrm{d}) \;,
\end{equation}
with the Heaviside step function $\mathcal{H}$. A peak is then located
at the 3-d position $(\vec{\theta}_0, z_\mathrm{d})$ that maximises
the likelihood $\mathcal{L}$. For the purpose of this work $10$
equally spaced redshift steps $z_\mathrm{d} = 0.1 \ldots 1.0$ were
used. The background galaxies were put into redshift bins with width
$\Delta z = 0.01$ assuming perfect knowledge of their redshifts.

As in \citet{2007A&A...470..821D} we used the weight function proposed
by \citet{2007A&A...462..875S}
\begin{equation}
  \label{eq:9}
  Q_\mathrm{NFW}(x; x_\mathrm{c}) \propto \frac{1}{1 + \ex^{6-150x} +
    \ex^{-47+50x}} \frac{\tanh(x / x_\mathrm{c})}{x / x_\mathrm{c}}\;,
\end{equation}
where $x = \vartheta / \vartheta_\mathrm{max}$ and $x_\mathrm{c}$ is a
free parameter, which was fixed to the value of $x_\mathrm{c} = 0.15$
determined to be ideal for the detection of galaxy clusters by
\citet{2005A&A...442...43H}. $Q_\mathrm{NFW}$ follows the shear
profile of an NFW halo with exponential cut-offs as $x \rightarrow 0$
or $x \rightarrow \infty$. 

The absolute scale $\vartheta_\mathrm{max}$ determines the halo radius
or mass to which the filter function is tuned. The filter scale chosen
for our simulations is $5\farcm6$ on the sky, corresponding to a
radius of $2\hm\,$\,Mpc at a redshift of $z = 0.3$. At this redshift
the lensing efficiency of our survey is maximal and the chosen radius
is adjusted to cluster sizes easily detectable with weak lensing while
smoothing over smaller halos. This smoothing also ensures that shot
noise from unresolved structures in the $N$-body simulations does not
play a role.

Peaks were detected by connected-component labelling of pixels above a
detection threshold. We used the 8 connectivity in 2-d and the 26
connectivity in 3-d, i.e., we consider all pixels that are connected
via the sides, edges, or corners of a square or a cube as one
structure. Additionally, for tomographic peaks the condition was
imposed that peaks must be detected in at least three adjacent
redshift bins. This additional requirement is used to filter out
detections at very high or low redshifts whose true redshift is
outside the tomography cube. Such peaks would pile up in the lowest
and highest redshift bin and lead to high additional noise in them. At
the same time this filter criterion suppresses the inclusion of peaks
caused by increasing shot noise at high redshifts caused by the
sharply decreasing number density of background galaxies. Such peaks
occur typically in only one or two redshift slices.

\subsection{Analysis}
\label{sec:analysis}
For every cosmological model the peak function gives a $p$-dimensional
data vector $\vec{\zeta}$ of observables. We will explore several
choices of observables below. The choice of cosmological parameters is
denoted by $\vec{\pi}$ and the model prediction is
$\vec{m}(\vec{\pi})$. The posterior probability distribution is
\begin{equation}
  \label{eq:10}
  p(\vec{\pi} | \vec{\zeta}) = \frac{p(\vec{\zeta} |
    \vec{\pi})}{p(\vec{\zeta})} p(\vec{\pi})\;,
\end{equation}
where $p(\vec{\pi})$ is the prior probability distribution,
$p(\vec{\zeta} | \vec{\pi})$ is the likelihood, and $p(\vec{\zeta})$
is the evidence. We used a flat prior with cutoffs, i.e.,
$p(\vec{\pi}) = 1$ if $\Omega_\mathrm{m} \in [0.1:0.5]$ and $\sigma_8
\in [0.4:1.1]$ and $p(\vec{\pi}) = 0$ otherwise. The evidence in our
case simply is a normalisation of the posterior obtained by
integrating the likelihood over the support of the prior.

Assuming that $\vec{\zeta}$ has a Gaussian distribution, the likelihood is
\begin{equation}
  \label{eq:11}
  \begin{split}
    p(\vec{\zeta} | \vec{\pi}) = & \frac{1}{(2\pi)^{d/2} \sqrt{\det
        \mathsf{\Sigma}(\vec{\pi})}} \\
    & \times \exp\left\{ -\frac{1}{2} \left[ \vec{\zeta} - m(\vec{\pi})
      \right]^\mathrm{t} \mathsf{\Sigma}^{-1}(\vec{\pi}) \left[ \vec{\zeta} -
        m(\vec{\pi}) \right] \right\}\;,
  \end{split}
\end{equation}
where $\mathsf{\Sigma}(\vec{\pi})$ is the covariance matrix of the
$d$-dimensional vector $\vec{\zeta}$. Since our parameter space is
covered only by discrete points we will compute $\vec{\zeta}$ by
fitting smooth functions to our data vectors or by interpolating
across our parameter space. Details will be given in
Sect.~\ref{sec:results}.

We obtained estimates $\hat{\mathsf{C}}$ of the covariance matrices
for the shear tomography and the peak statistics, as well as their
cross-covariance, from the field-to-field variation in the 175
ray-tracing simulations for the fiducial cosmology. As indicated in
Eq.~(\ref{eq:11}), the covariance in principle depends on cosmology.
Since we do not have a sufficient number of simulations for other
cosmological parameters, we set
$\mathsf{\Sigma}(\vec{\pi})=\mathsf{\Sigma}(\vec{\pi}_0)$. Although
this is an approximation commonly made, neglecting the cosmology
dependence of $\mathsf{\Sigma}$ can have a non-negligible impact on
the shape of the posterior likelihood, as has been investigated in
\citet{2008A&A...482....9E} for the case of cosmic shear. Furthermore,
we note that the assumption of a Gaussian likelihood is not
necessarily justified \citep{hartlap09,schneider09}.  These studies
suggest that both approximations lead to an over-estimation of the
errors on the cosmological parameters, so that our confidence regions
are most likely very conservative.

A further issue to keep in mind when estimating a covariance matrix
from the data or from simulations is that, although the covariance
$\hat{\mathsf{C}}$ estimated from the data is an unbiased estimator of
the true covariance $\mathsf{\Sigma}$, the inverse
$\hat{\mathsf{C}}^{-1}$ is not an unbiased estimator of
$\mathsf{\Sigma}^{-1}$. For $n$ independent simulations an unbiased
estimator of the inverse covariance is \citep{2007A&A...464..399H}
\begin{equation}
  \label{eq:12}
  \mathsf{\widehat{C^{-1}}} = \frac{n - d - 2}{n - 1}
  \hat{\mathsf{C}}^{-1}\; ,
\end{equation}
which is what we used when computing Eq.~(\ref{eq:11}). However, the
estimated covariance matrix becomes singular if $d > n - 1$,
which means that the limited number of ray-tracing simulations
available to us constrains the number of bins that can be used for the
analysis.

\section{Results}
\label{sec:results}
In analogy to the mass function $N(M, z | \vec{\pi})$, the peak
function measures the abundance of peaks as a function of convergence
and redshift $N(\kappa, z | \vec{\pi})$, where for single structures
along the line-of-sight the variation of the conversion from $\kappa$
to $M$ with redshift is given by the kernel~(\ref{eq:8}). Because we
detected peaks not in convergence maps but in aperture-mass maps or
cubes, the SNR of peaks was used as a proxy for mass.

\subsection{Constraints from aperture mass maps}
\label{sec:constr-from-apert}
As a first step we show that constraints on $\vec{\pi} =
(\Omega_\mathrm{m}, \sigma_8)^\mathrm{t}$ can be obtained from the
peak statistics in the absence of redshift information. A similar
study was recently presented by \citet{2009ApJ...691..547W}, who
demonstrated that parameters of the Dark Energy equation of state can
be constrained from high convergence regions. We detected peaks in
aperture-mass SNR maps as described in Sect.~\ref{sec:analysis} with a
detection threshold of $3.25\sigma$.

\begin{figure}
  \resizebox{\hsize}{!}{\includegraphics[clip]{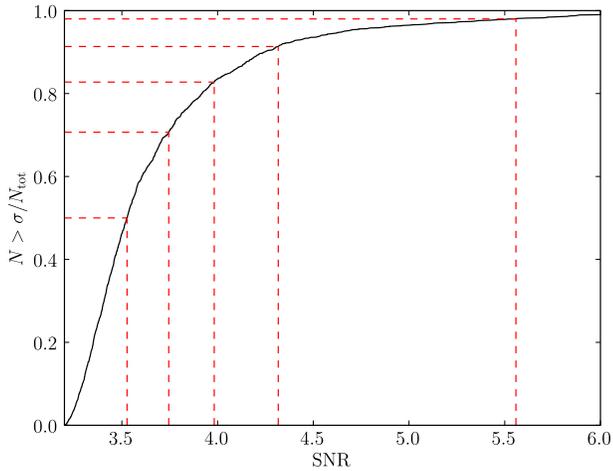}}
  \caption{Construction of the function $\vec{S}$. The solid black
    line is the cumulative SNR distribution of peaks detected in one
    of our 35 realizations of the fiducial cosmology. The horizontal
    dashed lines are the logarithmically spaced percentiles from
    $f_\mathrm{min} = 0.5$ to $f_\mathrm{max} = 0.98$ at which the
    cumulative SNR distribution is sampled. The corresponding SNR
    values denoted by the vertical dashed lines are the values in our
    data vector.}
  \label{fig:s-function}
\end{figure}

Binning the peaks by SNR is not an ideal way to use information about
their projected mass since either high SNR bins in cosmologies with
low clustering remain empty, or very massive peaks are beyond the SNR
of the maximum bin. Instead we used the cumulative SNR distribution of
peaks. The function $\vec{S}(\Omega_\mathrm{m}, \sigma_8): \mathbb{R}^2 \rightarrow
\mathbb{R}^{n_\mathrm{bin}}$ gives the SNR at which the cumulative
distribution exceeds the $f$th percentile for $n_\mathrm{bin}$ values
of $f$ ranging from $f_\mathrm{min}$ to $f_\mathrm{max}$.
Figure~\ref{fig:s-function} illustrates how $\vec{S}$ is constructed.

We measured $\vec{S}(\Omega_\mathrm{m}, \sigma_8)$ for $n_\mathrm{bin} = 5$
logarithmically spaced values from $f_\mathrm{min} = 0.50$ to
$f_\mathrm{max} = 0.98$. At the fiducial cosmology these percentiles
corresponds to SNR values of $3.5\sigma$ and $5.7\sigma$,
respectively. Typically several hundred peaks per $36$\,sq.\,deg.
field were detected so that the $98\%$ile could be reliably measured.

We used bilinear smoothing splines \citep{1993csfw.book.....D} to
interpolate $\vec{S}(\Omega_\mathrm{m}, \sigma_8)$ on the grid covered by our $N$-body
simulations. In this section splines are a sufficient description of
the variation of $\vec{S}$ over our parameter space because we only
seek to qualitatively demonstrate the ability of the peak statistics
to constrain cosmological parameters and to illustrate some of its
properties. We will use a more quantitative approach in the following
sections. 

\begin{figure}
  \resizebox{\hsize}{!}{\includegraphics[clip]{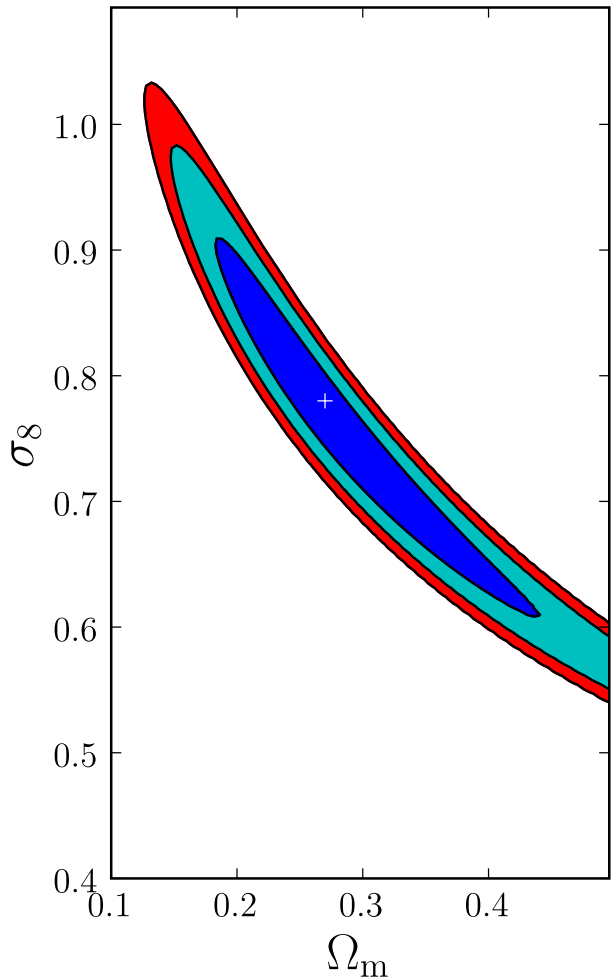}}
  \caption{Confidence contours of the aperture mass peak
    statistics. Shown are the $1$-, $2$-, and $3\sigma$ confidence
    contours of the $\vec{S}$ statistics. The white cross denotes the
    fiducial cosmology.}
  \label{fig:mapconf}
\end{figure}

Figure~\ref{fig:mapconf} shows the confidence contours derived from
this statistics in the $\Omega_\mathrm{m}$-$\sigma_8$ plane. They have
a shape similar to that seen in constraints derived from cluster
cosmology \citep[e.g.,][]{2009ApJ...691.1307H} and cosmic shear
\citep[e.g.,]{2008A&A...479....9F} for a CFHTLS like $180$\,sq.\,deg.
surey. In order to achieve this, we scaled the covariance, which we
computed for the individual $36$\,sq.\,deg.  fields back to the full
survey.  The similarity of the constraints is of course no surprise
since the peak statistics measures the same density fluctuations as
clusters of galaxies and cosmic shear.

Although the spline interpolation is mostly illustrative, we defined a
figure of merit (FoM), in analogy to the FoM of the Dark Energy Task
Force \citep{2006astro.ph..9591A}, as the inverse of the area inside
the $95\%$ confidence contour. We used this FoM to characterize how
the peak statistics changes when parameters entering the function
$\vec{S}$ are modified. Here in particular we examined the dependence
of the cosmological constrains on the minimum significance of a
detection. 

The detection threshold employed in the production of
Fig.~\ref{fig:mapconf} is very low and a sizable fraction of the peaks
detected in this way are simply due to shape noise
\citep{2007A&A...470..821D} and do not carry cosmological information.
However, at such a low detection threshold most peaks not caused by
noise fluctuations are also not due to a single massive halo but
caused by the alignment of LSS along the LOS. We demonstrate that
these low significance peaks indeed carry cosmological information by
comparing the FoM of the statistics in Fig.~\ref{fig:mapconf} to the
FoM resulting from the same function $\vec{S}$ with a detection
threshold of $4.5\sigma$. While the constraints in
Fig.~\ref{fig:mapconf} correspond to a FoM of $40$, the higher
detection threshold results in a FoM of only $20$. We note that the
$95\%$ confidence interval is not fully contained in the support of
our flat prior. For the low SNR detection, the $95\%$ confidence
interval is cut off by the prior only at the high
$\Omega_\mathrm{m}$/low $\sigma_8$ end. The prior terminates the
banana shaped confidence region at both ends for the high SNR
detection constraints. Consequently, the true figures of merit for
these statistics are smaller than presented here but more so for the
higher detection threshold, making the importance of low SNR peaks
even more evident. However, because we used a very simple
interpolation scheme, these numbers can only be rough guidelines and
we will present a more detailed assessment of the $\vec{S}$ function
in the next section.

We emphasize that, although there is no reason to believe that
the true values of $\Omega_\mathrm{m}$ or $\sigma_8$ are outside the
support of our prior, the prior is used only to avoid extrapolating
beyond the parameter range covered by the $N$-body simulations. Since
the aim of this study is to examine how well the peak statistics can
constrain cosmological parameters, we did not regard the prior as
information that should be included in the calculation of the FoM.

\subsection{Constraints from peak tomography}
\label{sec:constr-from-peak}
In this section we make use of the redshift information in our shear
catalogues. We employed the tomographic aperture mass outlined in
Sect.~\ref{sec:tomogr-apert-mass} to locate high convergence regions
not only in projection on the sky but also along the redshift axis.
With 175 independent ray-tracing simulations we could not compute the
covariance of the full peak function $\vec{N}(S, z)$ for meaningful
number of bins in SNR and redshift. Instead, we constructed two
separate peak functions from the tomographic data cubes. 

The first function measures the abundance of peaks in every redshift
bin as a function of cosmology only. We detected peaks as described in
Sect.~\ref{sec:tomogr-apert-mass} with a minimum detection threshold
of $\sigma_\mathrm{min} = 3.2$ and assigned them to a redshift bin
based on the redshift $z_\mathrm{d}$ that maximises the
likelihood~(\ref{eq:7}).  With 10 redshift bins, the vector-valued
function $\vec{M}(\Omega_\mathrm{m}, \sigma_8): \mathbb{R}^2
\rightarrow \mathbb{R}^{10}$ counts the number of peaks in each
redshift bin as a function of cosmological parameters. The second
function $\vec{S}$ uses the SNR information of the detected peaks as
defined in the previous section. We used the same values for
$f_\mathrm{min}$, $f_\mathrm{max}$, and $n_\mathrm{bin}$.

\begin{figure}
  \resizebox{\hsize}{!}{\includegraphics[clip]{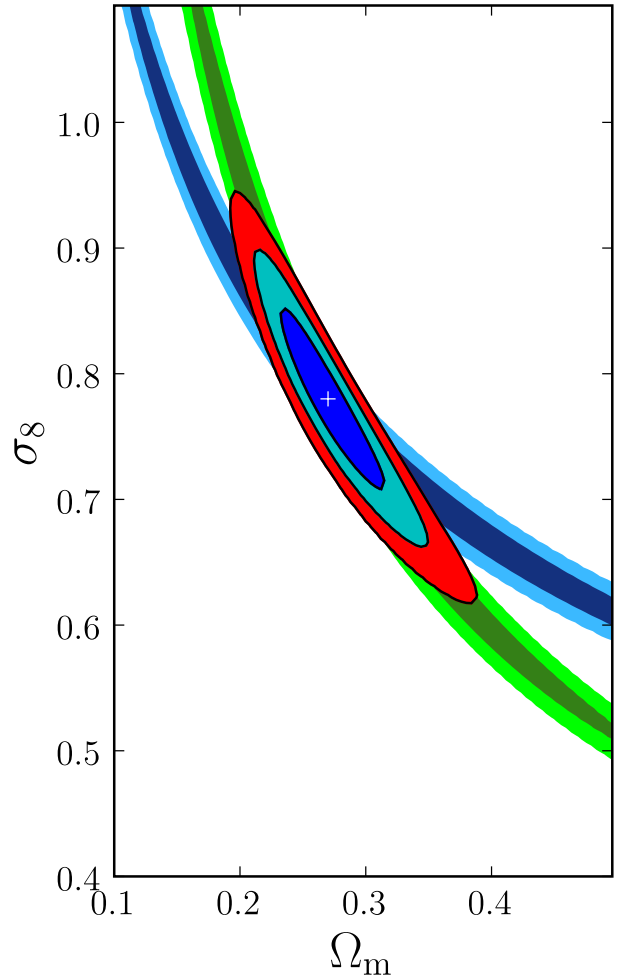}}
  \caption{Confidence contours of the peak statistics. Green and blue
    contours are $1$- and $2\sigma$ contours for the $\vec{M}$ and
    $\vec{S}$ statistics, respectively. Limited by black contour lines
    are the joint $1$-, $2$-, and $3\sigma$ confidence intervals.}
  \label{fig:peak-contours}
\end{figure}

To interpolate $\vec{M}(\Omega_\mathrm{m}, \sigma_8)$ and $\vec{S}(\Omega_\mathrm{m}, \sigma_8)$ between
points covered by our $N$-body simulations, we determined fitting
functions. These are given in Appendix~\ref{sec:fitt-funct-tomogr} and
are typically accurate to $\approx 10\%$.
Figure~\ref{fig:peak-contours} shows confidence contours in the
$\Omega_\mathrm{m}$-$\sigma_8$ plane obtained using these fitting
functions individually and for the combination of both peak functions,
$\vec{\zeta_\mathrm{p}} = (\vec{M}, \vec{S})^\mathrm{t}$. As one would
expect from cosmic shear and the cluster mass function, there is a
significant degeneracy between $\Omega_\mathrm{m}$ and $\sigma_8$. The
interesting result in Fig.~\ref{fig:peak-contours} is that $\vec{M}$
and $\vec{S}$ contain complementary information such that their
degeneracies are partly broken. Although the kernel~(\ref{eq:8}) is
very broad and the determination of a peak's redshift consequently
noisy \citep[see also][]{2005ApJ...624...59H} the information on the
growth of structure with redshift can be statistically recovered with
peak tomography. This then provides information not contained in
$\vec{S}$, which is equivalent to a projected mass function. It is
important to emphasise that we utilised the full cross-covariance
between $\vec{M}$ and $\vec{S}$ when we computed the joint confidence
contours.

We now revisited the issue of how the various parameters of the
$\vec{S}$ function change the information content of the peak
statistics. In the previous section we showed that the detection
threshold is an important parameter and that the number of projected
low SNR peaks helps to constrain parameters. The same is true for the
tomographic peak statistics. The FoM of the $\vec{S}$-function in
Fig.~\ref{fig:peak-contours} is $48$. Increasing the detection
threshold to $4\sigma$ and keeping all other parameters unchanged
decreases the FoM to only $7$. Like in the case of the aperture-mass
peak statistics, the confidence region is terminated by the prior and
the true difference between the different detection thresholds is even
larger than suggested by the FoM.

We note that the FoM of the $\vec{S}$ function for peak catalogues
generated from tomography cubes is not much higher than the rough
estimate of the FoM in Sect.~\ref{sec:constr-from-apert} of the
projected peak statistics. The $95\%$ confidence interval in the
tomographic case is not fully contained in the parameter space
explored by our simulations, whereas, with the spline approximation
from the previous section, the low $\Omega_\mathrm{m}$/high $\sigma_8$
end of the confidence contours is within our parameter range. In any
case, the substantial gain of the tomographic peak statistics does not
come from the deprojection of structures along the LOS but from
localising peaks along the redshift axis, i.e., from the combination
of the $\vec{S}$ and $\vec{M}$ statistics.

\begin{table}
  \caption{Figures of Merit of different parameters of the $\vec{S}$
    function.} 
  \label{tab:s-stat-FoM}
  \begin{tabular}{rrrrr}
    \hline\hline
    $f_\mathrm{min}$ & $f_\mathrm{max} $ & $n_\mathrm{bin}$ &
    $\sigma_\mathrm{min}$ & FoM  \\\hline
    0.50 & 0.98 &  5 & 3.2 & 48 \\
    0.50 & 0.98 &  5 & 4.0 &  7 \\
    0.08 & 0.98 & 10 & 3.2 & 48 \\
    0.08 & 0.50 &  5 & 3.2 & 34 \\
    \hline
  \end{tabular}
\end{table}

We also tested several choices of $f_\mathrm{min}$ and
$f_\mathrm{max}$ and the number of bins; Table~\ref{tab:s-stat-FoM}
gives an overview of various settings. Extending $f_\mathrm{min}$ to
lower values adds almost no information, even if the number of bins is
increased to preserve the information in the high SNR bins. For
example, decreasing $f_\mathrm{min} = 0.08$ and setting
$n_\mathrm{bin} = 10$ does not improve the FoM. The
information content of the $\vec{S}$ function is slightly higher
with these settings, as is evidenced by a $3\%$ decrease of the
area inside the $68\%$ confidence interval. However, the $95\%$
confidence contours within the support of our prior are not
tightened.

Despite of this, most of the information is contained in the low SNR
regime, as can be seen from the last two rows of
Table~\ref{tab:s-stat-FoM}. This information, however, can be
recovered with only one or two bins: Figure~\ref{fig:s-function} shows
that the low SNR end of the cumulative distribution function has an
almost constant slope. This slope is completely determined by the SNR
sampled at $f_\mathrm{min}$ and the point $(\sigma_\mathrm{min},0)$,
and explains why adding more points at the low end does not increase
the FoM. Further information comes only from the shape of the curve in
the intermediate SNR range. At the high SNR end the cumulative
distribution again has a constant slope and is fully characterized by
the last two sample points.

\subsection{Combination with cosmic shear}
\label{sec:comb-with-cosm}
Up to now cosmological information has generally been extracted from
lensing surveys via the cosmic-shear two-point correlation functions
(2PCF) \citep[e.g.,][]{2006sassfeebookS},
\begin{equation}
  \label{eq:13}
  \hat \xi_\pm(\theta) = \langle \epsilon_\mathrm{t}(\vec{\theta})
  \epsilon_\mathrm{t}(\vec{\theta} + \vec{\vartheta}) \rangle 
  \pm
  \langle \epsilon_\times(\vec{\theta}) \epsilon_\times(\vec{\theta} +
  \vec{\vartheta}) \rangle \;,
\end{equation}
which are related to the convergence power
spectrum $\mathcal{P}_\kappa$ via \citep[e.g.,][]{1992ApJ...388..272K}
\label{eq:14}
\begin{align}
  \xi_+(\theta) & = \int_0^\infty \frac{\dif l }{2\pi} J_0(l\theta)
  \mathcal{P}_\kappa(l) \\
  \xi_-(\theta) & = \int_0^\infty \frac{\dif l}{2\pi} J_4(l\theta)
  \mathcal{P}_\kappa(l)
\end{align}
Here $\epsilon_\times$ is the cross-component of the ellipticity and
the $J_n(x)$ are the $n$-th Bessel functions of the first kind.
Surveys using this method have led to increasingly tight constraints
in the $\Omega_\mathrm{m}$-$\sigma_8$ plane
\citep[e.g.,][]{2006ApJ...644...71J,2006A&A...452...51S,2006ApJ...647..116H,2007A&A...468..859H,2007MNRAS.381..702B,2008A&A...479....9F}.
However, the cosmic shear 2PCF describes the underlying density
fluctuations only completely if they are purely Gaussian. Cosmic shear
can access information about the non-Gaussianity of the matter
distribution only through higher-order correlation functions
\citep[see e.g.,][ for constraints using the three-point correlation
function]{2003MNRAS.340..580T}. The peak statistics on the other hand
is most sensitive to extreme overdensities along the LOS, i.e., to
those structures that contain most information about non-Gaussianity.
It is thus reasonable to assume that both statistics are not
completely degenerate and that combining the two does not simply
amount to using the same information twice. This expectation is
supported by a number of studies looking at the constraints one can
place on the Dark Energy equation of state by combining the cluster
mass function with other cosmological probes, including weak
gravitational lensing
\citep{2007PhRvD..75d3010F,2007NJPh....9..446T,2009arXiv0904.1589C}.
\citet{2007NJPh....9..446T} in particular examined the full
cross-covariance between the cosmic shear 2PCF and cluster counts of
shear-selected halos and found that the combination of both methods
leads to tighter constraints than either method alone can provide.

Cosmic shear, like the peak statistics, can greatly benefit from the
inclusion of redshift information by following the evolution of
structure with time \citep{1999ApJ...522L..21H,2005MNRAS.363..723B}.
This is done by dividing the galaxy sample into redshift bins and
computing their auto- and cross-correlation functions,
\begin{equation}
  \label{eq:15}
  \hat \xi^{(ij)}_\pm(\theta) = \langle \epsilon_\mathrm{t}^{(i)}(\vec{\theta})
  \epsilon^{(j)}_\mathrm{t}(\vec{\theta} + \vec{\vartheta}) \rangle 
  \pm
  \langle \epsilon^{(i)}_\times(\vec{\theta})
  \epsilon^{(j)}_\times(\vec{\theta} + \vec{\vartheta}) \rangle \;.
\end{equation}
We used this tomographic 2PCF to compare and combine the constraints
obtained from cosmic shear to those from the peak statistics in the
same survey.

We split the galaxy catalogue into two redshift bins, separated at
redshift $z = 0.6$ and estimated the tomographic cosmic shear 2PCF
$\xi^{(ij)}_\pm$ in our simulation of the fiducial cosmology for 60
logarithmically spaced intervals from $30\arcsec$ to $6\degr$. From
this we constructed data vectors by averaging the values of 6
consecutive bins into one entry in the data vector, so that we have a
60-dimensional cosmic shear data vector $\vec{\zeta_\mathrm{cs}} =
(\vec{\hat{\xi}^{(11)}_+},
\vec{\hat{\xi}^{(11)}_-},\vec{\hat{\xi}^{(12)}_+},
\vec{\hat{\xi}^{(12)}_-},\vec{\hat{\xi}^{(22)}_+},
\vec{\hat{\xi}^{(22)}_-})^\mathrm{t}$. Using these vectors we computed
the covariance of our cosmic shear measurements at the fiducial
cosmology. The choice of redshift and spatial bins was motivated by the
limited number of independent realisations of our fiducial cosmology,
which limits the dimensionality of the data vector.

\begin{figure}
  \resizebox{\hsize}{!}{\includegraphics[clip]{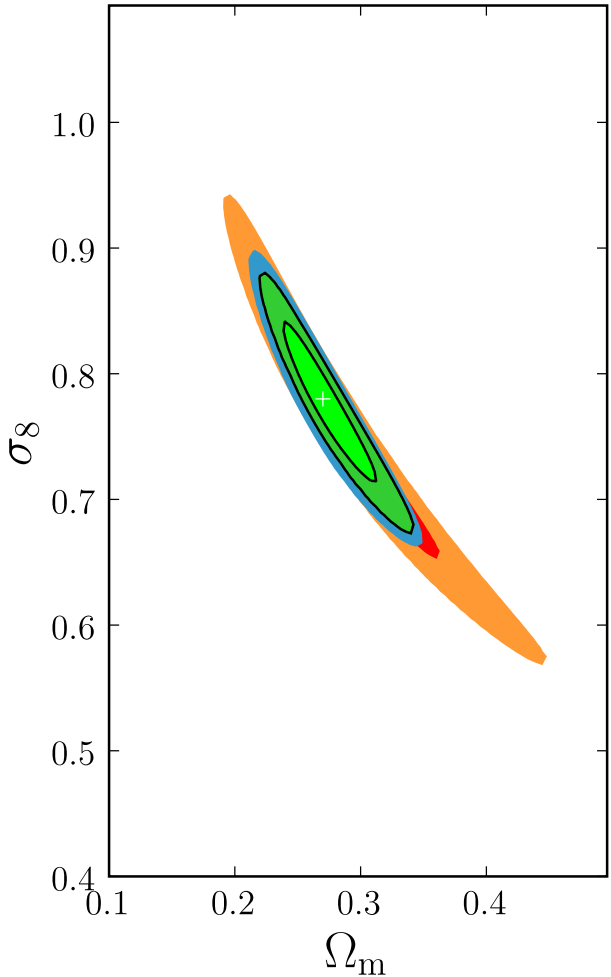}}
  \caption{Comparison of the cosmic-shear tomography confidence
    intervals (orange/red) with the full peak information making use of the
    combined $\vec{M}$ and $\vec{S}$ statistics (blue). Shown in green
    with black outlines is the combination of cosmic-shear and peak
    statistics. Again, $1$- and $2\sigma$ intervals are shown in all
    cases.}
  \label{fig:shear-peak-comb}
\end{figure}

\begin{table}
  \caption{Cosmological constraints using different statistics.}
  \label{tab:constraints}
  \begin{tabular}{lrrr}
    \hline\hline
    Type & \multicolumn{1}{c}{$\Omega_\mathrm{m}$} & 
    \multicolumn{1}{c}{$\sigma_8$} & FoM \\\hline
    Cosmic shear & $0.291^{+0.117}_{-0.091}$ &
    $0.756^{+0.155}_{-0.160}$ & 71 \\
    Peak statistics & $0.273^{+0.063}_{-0.053}$ &
    $0.776^{+0.107}_{-0.096}$ & 123 \\ 
    Combined &  $0.275^{+0.057}_{-0.051}$ &
    $0.774^{+0.095}_{-0.087}$ & 173 \\
    \hline
  \end{tabular}
\end{table}

We predicted the cosmic shear signal on a grid in our parameter space
using the transfer function of \citet{1998ApJ...496..605E} and the
non-linear power spectrum of \citet{1996MNRAS.280L..19P}.
Figure~\ref{fig:shear-peak-comb} shows a comparison of constraints
obtained using the cosmic shear 2PCF and the he $\vec{M}$ and
$\vec{S}$ peak statistics. The confidence region of the full peak
statistics $\zeta_\mathrm{p}$ is well aligned with the confidence
region of cosmic shear tomography. However, as
Table~\ref{tab:constraints} shows, the combined statistics
$\vec{\zeta} = (\vec{\zeta_\mathrm{p}}^\mathrm{t},
\vec{\hat{\xi}_\mathrm{cs}}^\mathrm{t})^\mathrm{t}$ still gives
significantly improved constraints; it has a FoM that is about $40\%$
larger than that of cosmic shear tomography alone.

\subsection{Stability of the constraints}
\label{sec:stab-constr}
As mentioned in Sect.~\ref{sec:analysis}, the estimated inverse
covariance becomes singular if $d > n - 1$, and the length of the data
vectors is consequentially limited by the number of ray-tracing
simulations at the fiducial cosmology. The combined statistics vector
$\vec{\zeta}$ is 75-dimensional, which is comparable to the number of
simulations $n = 175$.  Even though we can obtain an unbiased estimate
of the inverse covariance, this estimate is potentially very noisy.

\begin{figure}
  \resizebox{\hsize}{!}{\includegraphics[clip]{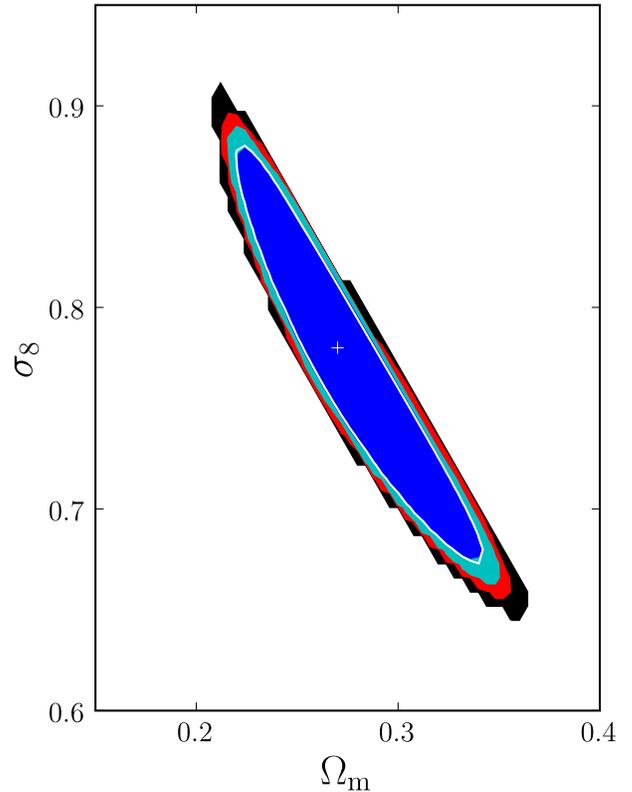}}  
  \caption{Variation of the constraints when bootstrapping the
    covariance matrix. The white contour line shows the $95\%$
    confidence interval of the combined peak and cosmic shear
    statistics from Fig.~\ref{fig:shear-peak-comb}. Shaded areas mark
    the regions inside the bootstrapped $95\%$ confidence interval for
    $x$ percent of the resamplings, where $x$ is $100$ (black), $99$
    (red), cyan(95), and blue (68).}
  \label{fig:stability}
\end{figure}

We estimate the effect of noise due to the finite number simulations
on the constraints derived in the previous section by we creating
$1000$ bootstrap-like samples from our set of simulations for the
fiducial cosmology. Each sample is constructed by randomly drawing
$n=175$ ray-tracing realisations with replacement. The covariance
matrix is estimated for each of the samples.  In doing so we kept
track how many independent simulations $n$ entered the estimation of
the inverse covariance matrix in Eq.~(\ref{eq:12}). At the same time
we ensured that we had enough independent simulations to estimate
$\mathsf{C^{-1}}$. Finally, we computed the corresponding confidence
regions as described before.  While not statistically rigorous, this
scheme nevertheless illustrates that the confidence intervals are
stable with respect to the set of simulations used.
Figure~\ref{fig:stability} shows the variation of the area inside this
confidence interval and illustrates that the dependence on the
simulations entering the computation of the covariance matrix is small
compared to the size of the confidence region.

We have also compared the confidence regions obtained using the
fitting functions of Appendix~\ref{sec:fitt-funct-tomogr} for the
function $\vec M$ with the constraints computed using the
interpolation method described in Section \ref{sec:constr-from-apert}.
While interpolating between simulations for different cosmologies is
noisy, the results using the fitting function might be affected by
accuracy and choice of the fitting functions. However, we do not find
significant differences between the two methods.

\section{Summary and Discussion}
\label{sec:discussion}
We showed in this paper that the number and properties of peaks found
in a weak lensing survey can constrain cosmological parameters. This
allowed us to turn an important limitation of weak-lensing cluster
searches, namely their low purity and completeness, into a source of
cosmological information. We note that a similar idea was recently
presented by \citet{2009ApJ...691..547W}. The most important
differences between their and our works are our purely numeric ansatz,
the inclusion of tomographic information, and the combination with
cosmic shear tomography.

In the pilot study presented here we demonstrated that the peak
statistics is able to provide constraints on $\Omega_\mathrm{m}$ and
$\sigma_8$, which are competitive with those obtained from cosmic
shear tomography on the same data set. By comparing the constraints
obtained from peaks found in maps of aperture mass and tomography
cubes we concluded that the tomographic redshift leads to much tighter
constraints than working with a 2-dimensional $\map$- or convergence
map alone. Even though the lensing efficiency varies only slowly with
lens redshift and the best estimate of a single peak's redshift has a
large scatter around the true redshift \citep{2005ApJ...624...59H},
the peak redshift distribution contains valuable information.

By looking at the SNR distribution function we showed that the SNR of
peaks, acting as a proxy for mass, also provides information beyond
the simple counting of peaks above a detection or mass threshold as it
was done by \citet{2007NJPh....9..446T} and
\citet{2007PhRvD..75d3010F}. We showed that peaks with a low SNR, most
of which cannot be associated with a single massive halo, contribute
significantly to the information content of the peak statistics. 

Finally, we investigated how the peak statistics compares with the
standard cosmic-shear tomography method and whether a combination of
both statistics is useful. We found that the peak statistics gives
constraints on $\Omega_\mathrm{m}$ and $\sigma_8$ that are competitive
with those expected from a cosmic-shear study on the same survey
fields. Taking at face value, our FoM suggests that the peak
statistics is indeed the superior method. The comparison, however, is
not entirely fair because we used only two redshift bins for the 2PCF
tomography while the peak statistics made use of almost perfect
redshift information, when we discretized the exact redshifts of the
background galaxies into bins of widths $\Delta z = 0.01$.  On the
other hand, \citet{2005ApJ...624...59H} found that even with very few
redshift bins, the tomographic peak finder is able to locate peaks
along the $z$-axis reasonably well and significantly outperforms a
simple $\map$ peak finder. We thus conclude that the peak statistics
gives constraints that are at least comparable to those obtainable
from more traditional weak lensing methods.

We combined the peak statistics with the cosmic-shear 2PCF tomography,
including the cross-covariance of both statistics measured at our
fiducial cosmology. Although both methods have a very similar
degeneracy, their combination improves our FoM by about $40\%$. We
surmised that this is due to the inclusion of information about the
non-Gaussianity of the matter density field in the peak statistics,
which is not contained in the 2PCF. This result is not unexpected.
\citet{2007NJPh....9..446T} already found that the combination of
cluster counts and shear tomography, including the full covariance,
improves cosmological constraints. The new information we add here is
that the gain in information continues to be present when cluster
counts are replaced with peak counts. The peak statistics, not
discriminating between massive halos and chance projections, acts as a
``very high-order'' cosmic shear method.

Eventually, the much more ambitious question one wants to answer is:
\emph{What is the ideal way to extract (lensing) information from present
and future cosmological surveys?} We made no attempt to tackle this
general problem but showed that a specific higher-order statistics, the
peak statistics we introduced in this paper, provides information that
can be combined with existing two-point statistics to improve
constraints. 

A great advantage of our numeric approach to the peak statistics is
that observational effects can be included by the simulator to an
almost arbitrary degree of precision, given of course these are known
and understood. An obvious example would be the inclusion of the holes
and gaps in survey fields caused in real data by bright stars,
diffraction spikes, ghost images, satellite tracks, and so on. 

The disadvantage of this numeric method is its enormous computational
cost. The generation of initial conditions, the $N$-body simulation,
ray-tracing, and the tomographic peak finder need about 150\,CPUh per
simulation on a cluster of Itanium Montecito 1.6\,Ghz CPUs. Most of
this time is used for the $N$-body simulation. Since the aperture mass
smoothes the convergence field with a kernel of typically $2$\,Mpc
radius high spatial resolution is not required. In this case the
computing time can be significantly reduced by replacing the TreePM
code with a simple particle-mesh algorithm. But even then, and in the
case of massive parallelisation, the wallclock time required to
simulate one point in parameter space would prohibit running a
standard Markov Chain Monte Carlo method. Population Monte Carlo
\citep[PMC,][]{2009arXiv0903.0837W} allows to investigate sample points
independently and thus facilitates the parallelisation beyond the
limits of effective interprocess communications of a single
simulation.

Another approach to reduce the computation time of Monte Carlo
simulations with $N$-body simulations is the framework
proposed by \citet{2007PhRvD..76h3503H}. They developed a mechanism
by which the parameter space of interest is optimally sampled with
relatively few high-precision simulations. Predictions for untried
positions are made from emulations rather than full simulations. An
important feature of this method is that the error bounds of the
emulations are constrained and that the uncertainties of the
emulator output are taken into account in the parameter estimation.
Either method, or the combination of PMC with emulators with
controlled error, bounds should make the application of the peak
statistics to current and future surveys feasible with current
high-performance computing hardware. Finally, we point out that the
growth of computing power expected from Moore's Law is faster than the
growth in etendue of envisioned survey facilities.

\section*{Acknowledgements}

This work was supported by AstroGrid-D, a project funded by the
Federal Ministry for Education and Research of Germany (BMBF) as part
of the D-Grid initiative. This research was in part supported by the
DFG cluster of excellence Origin and Structure of the Universe
(www.universe-cluster.de). JH acknowledges support by the Deutsche
Forschungsgemeinschaft within the Priority Programme 1177 under the
project SCHN 342/6, the BMBF through the TR33 ``The Dark Universe'',
and the Bonn-Cologne Graduate School of Physics and Astronomy. We are
very grateful to Torsten En{\ss}lin and Hans-Martin Adorf for their
support in getting us onto the grid. We thank Volker Springel for
making his initial conditions generator and friend-of-friend halo
finder available to us.

\bibliographystyle{aa}
\bibliography{FaLCoNS-I}

\begin{thebibliography}{50}
\expandafter\ifx\csname natexlab\endcsname\relax\def\natexlab#1{#1}\fi

\bibitem[{{Albrecht} {et~al.}(2006){Albrecht}, {Bernstein}, {Cahn}, {Freedman},
  {Hewitt}, {Hu}, {Huth}, {Kamionkowski}, {Kolb}, {Knox}, {Mather}, {Staggs},
  \& {Suntzeff}}]{2006astro.ph..9591A}
{Albrecht}, A., {Bernstein}, G., {Cahn}, R., {et~al.} 2006,
  arXiv:astro-ph/0609591

\bibitem[{{Bacon} {et~al.}(2005){Bacon}, {Taylor}, {Brown}, {Gray}, {Wolf},
  {Meisenheimer}, {Dye}, {Wisotzki}, {Borch}, \&
  {Kleinheinrich}}]{2005MNRAS.363..723B}
{Bacon}, D.~J., {Taylor}, A.~N., {Brown}, M.~L., {et~al.} 2005, \mnras, 363,
  723

\bibitem[{{Benjamin} {et~al.}(2007){Benjamin}, {Heymans}, {Semboloni}, {van
  Waerbeke}, {Hoekstra}, {Erben}, {Gladders}, {Hetterscheidt}, {Mellier}, \&
  {Yee}}]{2007MNRAS.381..702B}
{Benjamin}, J., {Heymans}, C., {Semboloni}, E., {et~al.} 2007, \mnras, 381, 702

\bibitem[{{Blandford} \& {Narayan}(1986)}]{1986ApJ...310..568B}
{Blandford}, R. \& {Narayan}, R. 1986, \apj, 310, 568

\bibitem[{{Cunha} {et~al.}(2009){Cunha}, {Huterer}, \&
  {Frieman}}]{2009arXiv0904.1589C}
{Cunha}, C., {Huterer}, D., \& {Frieman}, J.~A. 2009, arXiv:0904.1589

\bibitem[{{Dierckx}(1993)}]{1993csfw.book.....D}
{Dierckx}, P. 1993, {Curve and surface fitting with splines} (Oxford University
  Press)

\bibitem[{{Dietrich} {et~al.}(2007){Dietrich}, {Erben}, {Lamer}, {Schneider},
  {Schwope}, {Hartlap}, \& {Maturi}}]{2007A&A...470..821D}
{Dietrich}, J.~P., {Erben}, T., {Lamer}, G., {et~al.} 2007, \aap, 470, 821

\bibitem[{{Eifler} {et~al.}(2008){Eifler}, {Kilbinger}, \&
  {Schneider}}]{2008A&A...482....9E}
{Eifler}, T., {Kilbinger}, M., \& {Schneider}, P. 2008, \aap, 482, 9

\bibitem[{{Eisenstein} \& {Hu}(1998)}]{1998ApJ...496..605E}
{Eisenstein}, D.~J. \& {Hu}, W. 1998, \apj, 496, 605

\bibitem[{{Fang} \& {Haiman}(2007)}]{2007PhRvD..75d3010F}
{Fang}, W. \& {Haiman}, Z. 2007, \prd, 75, 043010

\bibitem[{{Fu} {et~al.}(2008){Fu}, {Semboloni}, {Hoekstra}, {Kilbinger}, {van
  Waerbeke}, {Tereno}, {Mellier}, {Heymans}, {Coupon}, {Benabed}, {Benjamin},
  {Bertin}, {Dor{\'e}}, {Hudson}, {Ilbert}, {Maoli}, {Marmo}, {McCracken}, \&
  {M{\'e}nard}}]{2008A&A...479....9F}
{Fu}, L., {Semboloni}, E., {Hoekstra}, H., {et~al.} 2008, \aap, 479, 9

\bibitem[{{Gavazzi} \& {Soucail}(2007)}]{2007A&A...462..459G}
{Gavazzi}, R. \& {Soucail}, G. 2007, \aap, 462, 459

\bibitem[{{Habib} {et~al.}(2007){Habib}, {Heitmann}, {Higdon}, {Nakhleh}, \&
  {Williams}}]{2007PhRvD..76h3503H}
{Habib}, S., {Heitmann}, K., {Higdon}, D., {Nakhleh}, C., \& {Williams}, B.
  2007, \prd, 76, 083503

\bibitem[{{Haiman} {et~al.}(2001){Haiman}, {Mohr}, \&
  {Holder}}]{2001ApJ...553..545H}
{Haiman}, Z., {Mohr}, J.~J., \& {Holder}, G.~P. 2001, \apj, 553, 545

\bibitem[{{Hamana} {et~al.}(2004){Hamana}, {Takada}, \&
  {Yoshida}}]{2004MNRAS.350..893H}
{Hamana}, T., {Takada}, M., \& {Yoshida}, N. 2004, \mnras, 350, 893

\bibitem[{{Hartlap} {et~al.}(2009){Hartlap}, {Schrabback}, {Simon}, \&
  {Schneider}}]{hartlap09}
{Hartlap}, J., {Schrabback}, T., {Simon}, P., \& {Schneider}, P. 2009,
  astro-ph/0901.3269

\bibitem[{{Hartlap} {et~al.}(2007){Hartlap}, {Simon}, \&
  {Schneider}}]{2007A&A...464..399H}
{Hartlap}, J., {Simon}, P., \& {Schneider}, P. 2007, \aap, 464, 399

\bibitem[{{Hennawi} \& {Spergel}(2005)}]{2005ApJ...624...59H}
{Hennawi}, J.~F. \& {Spergel}, D.~N. 2005, \apj, 624, 59

\bibitem[{{Henry} {et~al.}(2009){Henry}, {Evrard}, {Hoekstra}, {Babul}, \&
  {Mahdavi}}]{2009ApJ...691.1307H}
{Henry}, J.~P., {Evrard}, A.~E., {Hoekstra}, H., {Babul}, A., \& {Mahdavi}, A.
  2009, \apj, 691, 1307

\bibitem[{{Hetterscheidt} {et~al.}(2005){Hetterscheidt}, {Erben}, {Schneider},
  {Maoli}, {van Waerbeke}, \& {Mellier}}]{2005A&A...442...43H}
{Hetterscheidt}, M., {Erben}, T., {Schneider}, P., {et~al.} 2005, \aap, 442, 43

\bibitem[{{Hetterscheidt} {et~al.}(2007){Hetterscheidt}, {Simon}, {Schirmer},
  {Hildebrandt}, {Schrabback}, {Erben}, \& {Schneider}}]{2007A&A...468..859H}
{Hetterscheidt}, M., {Simon}, P., {Schirmer}, M., {et~al.} 2007, \aap, 468, 859

\bibitem[{{Hilbert} {et~al.}(2009){Hilbert}, {Hartlap}, {White}, \&
  {Schneider}}]{2009A&A...499...31H}
{Hilbert}, S., {Hartlap}, J., {White}, S.~D.~M., \& {Schneider}, P. 2009, \aap,
  499, 31

\bibitem[{{Hoekstra} {et~al.}(2006){Hoekstra}, {Mellier}, {van Waerbeke},
  {Semboloni}, {Fu}, {Hudson}, {Parker}, {Tereno}, \&
  {Benabed}}]{2006ApJ...647..116H}
{Hoekstra}, H., {Mellier}, Y., {van Waerbeke}, L., {et~al.} 2006, \apj, 647,
  116

\bibitem[{{Hu}(1999)}]{1999ApJ...522L..21H}
{Hu}, W. 1999, \apjl, 522, L21

\bibitem[{{Jain} {et~al.}(2000){Jain}, {Seljak}, \&
  {White}}]{2000ApJ...530..547J}
{Jain}, B., {Seljak}, U., \& {White}, S. 2000, \apj, 530, 547

\bibitem[{{Jarvis} {et~al.}(2006){Jarvis}, {Jain}, {Bernstein}, \&
  {Dolney}}]{2006ApJ...644...71J}
{Jarvis}, M., {Jain}, B., {Bernstein}, G., \& {Dolney}, D. 2006, \apj, 644, 71

\bibitem[{{Jenkins} {et~al.}(2001){Jenkins}, {Frenk}, {White}, {Colberg},
  {Cole}, {Evrard}, {Couchman}, \& {Yoshida}}]{2001MNRAS.321..372J}
{Jenkins}, A., {Frenk}, C.~S., {White}, S.~D.~M., {et~al.} 2001, \mnras, 321,
  372

\bibitem[{{Kaiser}(1992)}]{1992ApJ...388..272K}
{Kaiser}, N. 1992, \apj, 388, 272

\bibitem[{{Marian} \& {Bernstein}(2006)}]{2006PhRvD..73l3525M}
{Marian}, L. \& {Bernstein}, G.~M. 2006, \prd, 73, 123525

\bibitem[{{Marian} {et~al.}(2009){Marian}, {Smith}, \&
  {Bernstein}}]{2009ApJ...698L..33M}
{Marian}, L., {Smith}, R.~E., \& {Bernstein}, G.~M. 2009, \apjl, 698, L33

\bibitem[{{Miyazaki} {et~al.}(2007){Miyazaki}, {Hamana}, {Ellis}, {Kashikawa},
  {Massey}, {Taylor}, \& {Refregier}}]{2007ApJ...669..714M}
{Miyazaki}, S., {Hamana}, T., {Ellis}, R.~S., {et~al.} 2007, \apj, 669, 714

\bibitem[{{Peacock} \& {Dodds}(1996)}]{1996MNRAS.280L..19P}
{Peacock}, J.~A. \& {Dodds}, S.~J. 1996, \mnras, 280, L19

\bibitem[{{Press} \& {Schechter}(1974)}]{1974ApJ...187..425P}
{Press}, W.~H. \& {Schechter}, P. 1974, \apj, 187, 425

\bibitem[{{Schirmer} {et~al.}(2007){Schirmer}, {Erben}, {Hetterscheidt}, \&
  {Schneider}}]{2007A&A...462..875S}
{Schirmer}, M., {Erben}, T., {Hetterscheidt}, M., \& {Schneider}, P. 2007,
  \aap, 462, 875

\bibitem[{{Schneider}(1996)}]{1996MNRAS.283..837S}
{Schneider}, P. 1996, \mnras, 283, 837

\bibitem[{{Schneider}(2006)}]{2006sassfeebookS}
{Schneider}, P. 2006, in Saas Fee Advanced Course 33: Gravitational Lensing:
  Strong, Weak and Micro, ed. P.~{Schneider}, C.~{Kochanek}, \& J.~{Wambsganss}

\bibitem[{{Schneider} {et~al.}(1992){Schneider}, {Ehlers}, \&
  {Falco}}]{1992grle.book.....S}
{Schneider}, P., {Ehlers}, J.~., \& {Falco}, E.~E. 1992, {Gravitational Lenses}
  (Gravitational Lenses, XIV, 560 pp.~112 figs..~Springer-Verlag Berlin
  Heidelberg New York.~ Also Astronomy and Astrophysics Library)

\bibitem[{{Schneider} \& {Hartlap}(2009)}]{schneider09}
{Schneider}, P. \& {Hartlap}, J. 2009, astro-ph/0905.0577

\bibitem[{{Schneider} {et~al.}(1998){Schneider}, {van Waerbeke}, {Jain}, \&
  {Kruse}}]{1998MNRAS.296..873S}
{Schneider}, P., {van Waerbeke}, L., {Jain}, B., \& {Kruse}, G. 1998, \mnras,
  296, 873

\bibitem[{{Seitz} {et~al.}(1994){Seitz}, {Schneider}, \&
  {Ehlers}}]{1994CQGra..11.2345S}
{Seitz}, S., {Schneider}, P., \& {Ehlers}, J. 1994, Classical and Quantum
  Gravity, 11, 2345

\bibitem[{{Semboloni} {et~al.}(2006){Semboloni}, {Mellier}, {van Waerbeke},
  {Hoekstra}, {Tereno}, {Benabed}, {Gwyn}, {Fu}, {Hudson}, {Maoli}, \&
  {Parker}}]{2006A&A...452...51S}
{Semboloni}, E., {Mellier}, Y., {van Waerbeke}, L., {et~al.} 2006, \aap, 452,
  51

\bibitem[{{Sheth} \& {Tormen}(2002)}]{2002MNRAS.329...61S}
{Sheth}, R.~K. \& {Tormen}, G. 2002, \mnras, 329, 61

\bibitem[{{Smith} {et~al.}(2003){Smith}, {Peacock}, {Jenkins}, {White},
  {Frenk}, {Pearce}, {Thomas}, {Efstathiou}, \&
  {Couchman}}]{2003MNRAS.341.1311S}
{Smith}, R.~E., {Peacock}, J.~A., {Jenkins}, A., {et~al.} 2003, \mnras, 341,
  1311

\bibitem[{{Springel}(2005)}]{2005MNRAS.364.1105S}
{Springel}, V. 2005, \mnras, 364, 1105

\bibitem[{{Takada} \& {Bridle}(2007)}]{2007NJPh....9..446T}
{Takada}, M. \& {Bridle}, S. 2007, New Journal of Physics, 9, 446

\bibitem[{{Takada} \& {Jain}(2003)}]{2003MNRAS.340..580T}
{Takada}, M. \& {Jain}, B. 2003, \mnras, 340, 580

\bibitem[{{Wang} \& {Steinhardt}(1998)}]{1998ApJ...508..483W}
{Wang}, L. \& {Steinhardt}, P.~J. 1998, \apj, 508, 483

\bibitem[{{Wang} {et~al.}(2009){Wang}, {Haiman}, \&
  {May}}]{2009ApJ...691..547W}
{Wang}, S., {Haiman}, Z., \& {May}, M. 2009, \apj, 691, 547

\bibitem[{{Weller} {et~al.}(2002){Weller}, {Battye}, \&
  {Kneissl}}]{2002PhRvL..88w1301W}
{Weller}, J., {Battye}, R.~A., \& {Kneissl}, R. 2002, Physical Review Letters,
  88, 231301

\bibitem[{{Wraith} {et~al.}(2009){Wraith}, {Kilbinger}, {Benabed}, {Capp{\'e}},
  {Cardoso}, {Fort}, {Prunet}, \& {Robert}}]{2009arXiv0903.0837W}
{Wraith}, D., {Kilbinger}, M., {Benabed}, K., {et~al.} 2009, arXiv:0903.0837

\end{thebibliography}

\appendix

\section{Fitting functions for tomographic peaks}
\label{sec:fitt-funct-tomogr}
We find that each component of
$\vec{M}(\Omega_\mathrm{m},\sigma_8)$ is well described by the function
\begin{equation}
  \begin{split}
    \label{eq:16}
    M_z(\Omega_\mathrm{m},\sigma_8) = & A g(z) \left[1 + (1+z)^3\right]
    \Omega_\mathrm{m}^\beta \sigma_8^\alpha
    P(z, \Omega_\mathrm{m}, \sigma_8) \\
    & + p_3(\Omega_\mathrm{m}, \sigma_8)
  \end{split}
\end{equation}
where $g(z)$ is the distance ratio
$D_\mathrm{ds}/D_\mathrm{s}$ averaged over the source redshift
distribution \citep{1998MNRAS.296..873S}. The functions $P$ and
$p_3$ are polynomials, where
\begin{equation}
  \begin{split}
    \label{eq:17}
    p_1 &= p_{10} + p_{11} \Delta\Omega_\mathrm{m} + p_{12}
    \Delta\sigma_8 \\
    p_2 &= p_{20} + p_{21} \Delta\Omega_\mathrm{m} + p_{22}
    \Delta\sigma_8 \\
    p_3 &= p_{30} + p_{31} \Delta\Omega_\mathrm{m} + p_{32}
    \Delta\sigma_8 \\
    P &= 1 + z (p_1 + z p_2)\;,
  \end{split}
\end{equation}
with $\Delta\Omega_\mathrm{m} = \Omega_\mathrm{m} -
\Omega_\mathrm{m_0}$, $\Delta\sigma_8 = \sigma_8 - \sigma_{8_0}$. The
constant $A$, the
polynomial coefficients $p_{mn}$, and the exponents $\alpha$ and
$\beta$ are the free parameters of the fitting function, which are
fitted simultaneously for all redshift bins and cosmological parameters.
The values of the fit parameters depend on the choice of the
signal-to-noise-cut, the ellipticity dispersion, the redshift
distribution of the galaxies, the filter function and radius and the
galaxy number density. Quoting the best-fit values for our specific
choices of these parameters would therefore be of very limited use.

The SNR probability distribution $p(S; \Omega_\mathrm{m},\sigma_8)$ is well
described by a log-normal distribution for $x = \log(S^2/10)$
\begin{equation}
  \begin{split}
    \label{eq:18}
    p(x;\Omega_\mathrm{m},\sigma_8) = \frac{1}{\sqrt{2\pi} x S
      \sigma(\Omega_\mathrm{m},\sigma_8)} \times \\
    \exp\left\{
      -\frac{\left[\log(x) -
          \mu(\Omega_\mathrm{m},\sigma_8)\right]^2}{2\sigma^2(\Omega_\mathrm{m},\sigma_8)}\right\}
    \;,
  \end{split}
\end{equation}
where the cosmological dependence of $\mu$ and $\sigma$ are described
by (double) power laws
\begin{equation}
  \begin{split}
    \label{eq:19}
    \mu(\Omega_\mathrm{m}, \sigma_8) = & A_1 \Omega_\mathrm{m}^{a_1}
    \sigma_8^{b_1} +
    \Omega_\mathrm{m}^{c_1} \sigma_8^{d_1}\\
    \sigma(\Omega_\mathrm{m}, \sigma_8) = & A_2
    \Omega_\mathrm{m}^{a_2} \sigma_8^{b_2} \;.
  \end{split}
\end{equation}
The cosmology dependent number count of peaks is given by
\begin{equation}
  \label{eq:20}
  N(\Omega_\mathrm{m}, \sigma_8) = A_3 \Omega_\mathrm{m}^{a_3}
  \sigma_8^{b_3} + \Omega_\mathrm{m}^{c_3} \sigma_8^{d_3}\;.
\end{equation}
Note that we do not fit the probability density function (\ref{eq:18})
to the data, but the cumulative distribution function which we compute
numerically.

\label{lastpage}
\end{document}